\documentclass[12pt]{article}
\usepackage{amsmath,amsfonts,amssymb}
\usepackage{graphicx}
\usepackage{enumerate}

\numberwithin{equation}{section}

\begin{document}
\newcommand{\todo}[1]{{\em \small {#1}}\marginpar{$\Longleftarrow$}}
\newcommand{\labell}[1]{\label{#1}\qquad_{#1}} %{\label{#1}} %

\rightline{DCPT-07/27}
\vskip 1cm

\begin{center}
{\Large \bf Reversing Renormalization-Group Flows with AdS/CFT}
\end{center}
\vskip 1cm

\renewcommand{\thefootnote}{\fnsymbol{footnote}}
\centerline{\bf Donald Marolf\footnote{marolf@physics.ucsb.edu} and Simon
F. Ross\footnote{S.F.Ross@durham.ac.uk}}
\vskip .5cm
\centerline{ \it Department of Physics, University of California}
\centerline{ \it Santa Barbara, CA 93106, U.S.A.}
\vskip .5cm
\centerline{ \it Centre for Particle Theory, Department of
Mathematical Sciences}
\centerline{\it University of Durham, South Road, Durham DH1 3LE, U.K.}

\setcounter{footnote}{0}
\renewcommand{\thefootnote}{\arabic{footnote}}

\vskip 1cm
\begin{abstract}
  For scalar fields in AdS with masses slightly above the
  Breitenlohner-Freedman bound, appropriate non-local boundary
  conditions can define a unitary theory.  Such boundary conditions
  correspond to non-local deformations of the dual CFT, and generate a
  non-local renormalization-group flow.  Nevertheless, a bulk analysis
  suggests that certain such flows lead to {\it local} CFTs in the
  infra-red.  Since the flows are non-local, they can either increase
  or decrease the central charge of the CFT.  In fact, given any local
  renormalization-group flow within a certain general class which
  leads from a UV theory (CFT$_1$) to an IR theory (CFT$_2$), we show
  that one can find such a non-local flow in which the endpoints are
  interchanged: the non-local theory flows from CFT$_2$ in the IR to
  CFT$_1$ in the UV.  We work at large $N$, but the flows we consider
  involve quantum field effects in the bulk, corresponding to
  $1/N$ corrections in the dual theory.
\end{abstract}

\section{Introduction}

In the AdS/CFT correspondence, boundary conditions for bulk fields are
related to the specification of the dual CFT
\cite{Maldacena:1997re,Witten:1998qj,Gubser:1998bc,Aharony:1999ti}.
Changes in the boundary conditions will correspond to deformations of
the dual CFT Lagrangian. Deformations by relevant operators will break
the conformal symmetry, generating a renormalization group flow from
the original CFT in the UV. This renormalization group flow is
identified via the UV/IR relation with the radial evolution of the
bulk spacetime. A prototypical example of this identification
is~\cite{Freedman:1999gp}, where a spacetime which is asymptotically
AdS both as $r \to \infty$ and as $r \to 0$ is identified with a
renormalization group flow from a UV fixed point described by one CFT to
an IR fixed point described by a different CFT. The bulk spacetime has
different effective cosmological constants in the two asymptotic
limits; these are identified with the central charges of the
corresponding CFTs.

Bulk scalar fields in AdS$_{d+1}$ with mass in the range $-d^2/4 \leq
m^2 < -d^2/4 +1$ provide an interesting and more subtle example of this
correspondence.  As indicated by the work of Breitenlohner and
Freedman~\cite{Breitenlohner:1982bm,Breitenlohner:1982jf}, such scalar
fields admit a variety of possible boundary conditions.  In
particular, one may fix either the faster or slower falloff part of
the scalar field at infinity.

The two resulting bulk theories correspond to two different dual
CFTs, in which the field $\phi$ is dual to operators of dimensions
$\Delta_-$ and $\Delta_+ = d-\Delta_-$ respectively, where $d/2
\geq \Delta_- > d/2-1$. In~\cite{Witten:2001ua,Berkooz:2002ug}, it
was observed that a linear boundary condition, relating
the faster falloff part to the slower, corresponds to a
double-trace deformation, adding a term $f \mathcal{O}^2$ to the
Lagrangian of the CFT. Starting from the $\Delta_-$ CFT, this
is a relevant deformation, which will produce an
renormalization-group flow which is expected to end at the
$\Delta_+$ CFT in the IR; evidence for this picture has been
obtained in~\cite{Gubser:2002zh,Gubser:2002vv,Hartman:2006dy}.
Since a double-trace operator corresponds to a multiparticle
state, the double-trace deformations in the CFT have also been
related to worldsheet non-locality in the bulk string
theory~\cite{Aharony:2001pa,Aharony:2001dp}.

This case is more subtle because pure anti-de Sitter space, with zero
scalar field, remains a solution of the theory at the classical level
for arbitrary linear boundary conditions on the scalar field. The
existence of a non-trivial renormalization group flow is only detected
from the bulk point of view once we take into account the one-loop
back reaction of the scalar field, which produces a modification of
the geometry depending on the boundary
conditions~\cite{Gubser:2002zh}.

For the classical renormalization group
flows,~\cite{Girardello:1998pd,Freedman:1999gp} established a
c-theorem, demonstrating that so long as the bulk geometry satisfied a
suitable energy condition, the c-function of~\cite{Cardy:1988cw} would
be monotonically decreasing along the flow from the UV to the IR (see
also \cite{Alvarez:1998wr}). In the case of flows associated with
general linear boundary conditions for a scalar field, no such general
argument exists, but~\cite{Gubser:2002zh} showed that the c-function
decreases along the flow from the $\Delta_-$ CFT to the $\Delta_+$
CFT. Thus, it seemed that the dual supergravity description
automatically enforces a kind of c-theorem, agreeing with the
intuitive expectation that the effective number of massless degrees of
freedom decreases along the flow.

 In this paper, we will show that more general boundary conditions can
produce quite different behaviour. The boundary described above corresponded to local deformations of the dual CFT but,  from
the bulk point of view, the allowed class of boundary conditions for a
scalar field (those boundary conditions that preserve the symplectic
flux) is much broader. We will discuss a particular example of this broader class
which corresponds to a non-local
deformation in the dual CFT, and which produces a renormalization group flow
which moves in the opposite direction to the one previously studied;
as a consequence, the c-function increases as we flow to the IR. 
The bulk perspective thus indicates that, at least at large $N$, there are non-local deformations
of the CFT which can produce controlled, well-defined renormalization
group flows, and that these flows violate our usual intuition about the
behaviour of the c-function. (Since the deformation is non-local, it
is perhaps not really surprising that the c-function increases; the
usual intuition that the renormalization group flow corresponds to
integrating out degrees of freedom on short scales does not obviously
apply in this case.) It is also interesting that although the flow is
non-local, the IR fixed point is a local CFT. 

In the next section, we will review the constraints on allowed
boundary conditions to have a well-defined bulk theory, and observe
that these allow for non-local boundary conditions.   
In section~\ref{vace}, we consider a particular simple non-local boundary condition and show that the corresponding deformation of the CFT leads to a renormalization group 
flow from the $\Delta_+$ CFT to the
$\Delta_-$ CFT, reversing the usual flow direction.    We close with some discussion in section \ref{disc}. 

\section{Local and Non-local boundary conditions}
\label{scalar}

Because AdS$_{d+1}$ is not globally hyperbolic, boundary conditions
play an especially important role in both classical and quantum
dynamics.  The most obvious classical role is to ensure that the
initial data on a hypersurface has a unique evolution into the future.
Though one could take the viewpoint that no more is required for
classical physics, it is natural to also demand that the system have a
well-defined phase space formulation.  Such a structure is in any case
pre-requisite for either canonical or path-integral quantization of
the theory. We therefore restrict to boundary conditions for which the
symplectic structure is finite and the symplectic flux\footnote{The
symplectic flux for a scalar field is proportional to the Klein-Gordon
flux.  See e.g.  \cite{WaldThermo,Lee:1990nz}, for general comments on
symplectic structures and their role in quantization.}  through
infinity vanishes, so that the symplectic structure is conserved.

 We consider only the low-energy effective theory in the bulk,
which we take to be given by a local field theory including the
metric as a dynamical degree of freedom.  In fact, we assume
that it is sufficient for our purposes to approximate even this
theory by that of a linear scalar field propagating on a fixed
AdS$_{d+1}$ background.  This approximation can be justified (see
e.g., \cite{Henneaux:2006hk,Amsel:2006uf}) for scalar fields with
mass sufficiently close to the Breitenlohner-Freedman bound.  The
assumption of linearity makes the physics particularly transparent
and allows us to make definite statements about the quantum
theory.  In section \ref{vace} below, we will also assume that the
boundary conditions are linear. The more general case can then be
dealt with perturbatively. 

 Consider such  a scalar field with mass $m$ in the range
\begin{equation}
\label{mrange} -\frac{d^2}{4} +1 > m^2 >  - \frac{d^2}{4}.
\end{equation}
propagating on a fixed AdS$_{d+1}$ background,
\begin{equation}
ds^2 = g_{ab} dy^a dy^b =  r^2 (-dt^2 +  \sum_{i=1}^{d-1} dx_i^2)+ \frac{dr^2}{r^2},
\end{equation}
where we have  restricted attention to the Poincar\'e patch and  fixed the AdS length scale to $\ell=1$, so $\Lambda =
-\frac{1}{2} d(d-1)$. 

Since we are interested in boundary conditions, we first describe the
asymptotic behavior of the field. Suppose that our scalar is
associated with a potential $V(\phi)$ with squared mass $m^2 =
\frac{1}{2} V''(0)$. All solutions to the equations of motion take the
asymptotic form
\begin{equation}
\label{asympt}
\phi \rightarrow \frac{\alpha(x)}{r^{\lambda_-}} +
\frac{\beta(x)}{r^{\lambda_+}},
\end{equation}
where $x$ are coordinates on null infinity ($\partial {\cal M}$, also
known as the conformal boundary) and where 
\begin{equation}
\lambda_\pm = \frac{d}{2} \pm \frac{1}{2} \sqrt{d^2 + 4 m^2}.
\end{equation}
Note that (\ref{mrange}) implies
\begin{equation}
\label{lrange}
2 > \lambda_+ - \lambda_- > 0.
\end{equation}
 
 The mass range (\ref{mrange}) is precisely the range for which all
solutions (\ref{asympt}) are normalizable with respect to the
symplectic structure (see e.g. \cite{Klebanov:1999tb}), except that we have  excluded the special case $m^2 =-d^2/4$.   Thus, the
only constraint is the requirement that the symplectic flux through infinity
vanish.  For two vectors $\delta_1\phi, \delta_2\phi$ tangent to
the space of solutions, the usual symplectic flux through a region $R$
of null infinity is
\begin{equation}
\label{sflux} \omega_R(\delta_1 \phi, \delta_2 \phi) = (\lambda_+
- \lambda_-) \int_R \sqrt{\Omega} ( \delta_1 \alpha \delta_2 \beta
- \delta_1 \beta \delta_2 \alpha ),
\end{equation}
where $\sqrt{\Omega}$ is the volume element on the boundary. 

If our boundary condition is to force (\ref{sflux}) to vanish for all
regions $R$, then $\alpha$ must be an ultra-local function of $\beta$;
i.e., $\alpha(x)$ can depend only on $\beta(x)$ at a point, and cannot
depend on derivatives of $\beta$:
\begin{equation}
\label{sgen}
\alpha (x) = J_\alpha(x, \beta) \ \ \ {\rm or} \ \ \ \beta (x) =
J_\beta(x, \alpha) .
\end{equation}
Note that each expression  (\ref{sgen})  admits a potential
 $W_\alpha(x,\beta)$, $W_\beta(x,\alpha)$  such that
\begin{equation} 
 \frac{\partial W_\alpha }{\partial \beta}=
\label{spot} (\lambda_+ - \lambda_-) J_\alpha (x,\beta),  \ \ \  \frac{\partial W_\beta }{\partial \alpha (x)}=
- (\lambda_+ - \lambda_-) J_\beta (x,\alpha) ,
 \end{equation}
where the normalization factor $(\lambda_+ - \lambda_-) $ on the
right-hand side was chosen for later convenience.  One may further
show that all such boundary conditions remain valid when the scalar
field is coupled to gravity; see \cite{Amsel:2006uf} for a general
analysis and
\cite{Henneaux:2002wm,Henneaux:2004zi,Hertog:2004dr,Hertog:2004ns,Hertog:2005hm,Henneaux:2006hk}
for direct calculations. 

So far, our discussion is fairly standard. We would now like to point
out that while it would not correspond to our usual notion of a local
bulk theory, one  {\it could}  choose to require the integrated flux
(\ref{sflux}) to vanish only for a certain family of regions\footnote{A different class of generalizations may be obtained by adding extra boundary degrees of freedom, and thus extra boundary terms to the symplectic structure.  The study of bulk vector fields in \cite{Witten:2003ya} appears to be an example of this other sort of generalization.}  $R$.  For
example, consider a theory which associates a phase space only to spacelike surfaces $\Sigma$ which are asymptotically of the form $t = constant$, independent of the spatial coordinates $\vec x$ on the boundary.  In such a theory, the flux is required to vanish only through regions $R$ bounded by
constant $t$ surfaces.  The boundary condition $J_\alpha(x,\beta)$ may then be much more general: while we still require ultra-locality in $t$, one may allow the boundary condition to depend on spatial derivatives of $\beta$, or even to be a non-local function of $\beta$ on each constant $t$ surface.  To emphasize the distinction between the dependence on $t$ and $\vec x$, let us write such a more general boundary condition as $J_\alpha(t,\vec x,\beta]$.   

At each $t$, we may allow $J_\alpha(t,\vec x,\beta]$ to be any
functional of $\beta(\vec x)$ for which $\frac{\delta J_\alpha(t,\vec
  x,\beta]}{\delta \beta(\vec y)}$ is the kernel of a self-adjoint operator in the $L^2$ space
associated with the relevant constant $t$ hypersurface.  Such
self-adjointness will hold when $J_\alpha$ can again be expressed in
terms of a (perhaps non-local) potential ${\cal W}_\alpha(t,\beta]$
via
\begin{equation}
(\lambda_+ - \lambda_-) J_\alpha (t, \vec x, \beta] = \frac{1}{\sqrt{h}} \frac{\delta {\cal W}_\alpha(t,\beta]}{\delta \beta(\vec y)},
\end{equation}
where $\sqrt{h}$ is the volume element associated with the spatial directions.  Here the relation between $J_\alpha$ and the potential ${\cal W}_\alpha$ involves functional derivatives, instead of the partial derivatives of (\ref{spot}).
A useful set of examples are of the form
\begin{equation}
\label{exs}
{\cal W}_\alpha (t, \beta]= \frac{\epsilon}{2} (\lambda_+ - \lambda_-) \int_{t= const} \sqrt{h} \beta  F_{\gamma}(\nabla)  \beta,
\end{equation}
where $F_{\gamma}(\nabla)$ is a homogeneous function of order $\gamma$ in the {\it spatial} covariant derivatives.  In general, $F_\gamma(\nabla) \beta$ may be defined by decomposing $\beta$ into normal modes.   The corresponding boundary condition is just 
\begin{equation}
\label{gamma}
\alpha = \epsilon F_\gamma(\nabla) \beta.
\end{equation} 
Of course, a similar construction is possible for $J_\beta$ and ${\cal W}_\beta$.   For the examples (\ref{exs}) we have
\begin{equation}
\label{exsb}
{\cal W}_\beta (t, \alpha]= \frac{1}{2 \epsilon} (\lambda_+ - \lambda_-) \int_{t= const} \sqrt{h} \beta  [F_\gamma(\nabla)]^{-1} \beta.
\end{equation}
One notes that $[F_\gamma(\nabla)]^{-1}$ is a homogeneous function of order $-\gamma$ and, as a result, at most one of ${\cal W}_\alpha$, ${\cal W}_\beta$ can be local for $\gamma \neq 0$.

 Let us now consider the effect of (\ref{gamma}) on the dynamics of bulk fields.  It is interesting to consider the evolution of initial data of compact support.  For short times, the boundary conditions will be irrelevant and the data will evolve in a purely local (and causal) fashion.  However, once the initial data propagates to the boundary of AdS space, the fields will feel the effects of the boundary condition.  For $\gamma \neq 0$, one expects that for at least one of $\alpha, \beta$,  the equations of motion will express 2nd time-derivatives as non-local functionals of the initial data. Thus, at this stage, the dynamics becomes non-local with respect to conformally compactified AdS (see figure 1).   When this is the case, we will say that the boundary condition is ``non-local," even if it may be possible to write the relation between $\alpha$ and $\beta$ in a local form (e.g., $\alpha = \nabla^2 \beta$ for $\gamma=2$).

\begin{figure}
\begin{center}
\includegraphics[width=2.5cm]{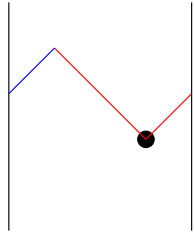} 
\end{center}
\caption{The effective causal structure of a theory with non-local boundary conditions.  The influence of a source (black dot) in the interior propagates causally toward the boundary.  However, once the signal reaches the boundary (right side), information can be instantaneously transferred elsewhere on the boundary (left side), and can then causally propagate back into the interior. }
\end{figure}

\section{One-loop vacuum energy and RG flow}
\label{vace}

 Let us now examine  the effects of the non-locality from the boundary point
of view. We first need to briefly review the relation between boundary
conditions for scalar fields and the associated deformations of the
dual field theory.  This correspondence was conjectured in
\cite{Witten:2001ua,Berkooz:2002ug}, derived in \cite{Sever:2002fk},
and studied further in, e.g.
\cite{Minces:2001zy,Muck:2002gm,Minces:2002wp,Minces:2004zr}. The
Lorentzian version of this analysis was discussed
in~\cite{Marolf:2006nd}, to which we refer the reader for further
detail. We  consider  only the case of masses strictly greater
than the Breitenlohner-Freedman bound, as only in this case is the UV
theory exactly conformal. There are two  particularly simple classes of boundary conditions,
fixing $\alpha$ or fixing $\beta$. These two boundary conditions
correspond to two different CFTs, coupled to sources of dimensions
$\Delta_\pm$.  

 If we consider the boundary conditions which fix
$\alpha$, so that $\alpha = J_\alpha(x)$ for some fixed function
$J_\alpha$ on $\partial \mathcal{M}$, then in the dual field theory
we can consider the operator dual to the source $J_\alpha$. In the (bulk) 
semi-classical limit its expectation value is given by
\begin{equation}
\label{scalarOa}
\langle {\cal O}_\alpha \rangle = \frac{1}{\sqrt{\Omega}} \frac{\delta S_{\alpha = const}}{\delta J_\alpha} = (\lambda_+ - \lambda_-) \beta.
\end{equation}
Similarly, in the theory defined by fixing $\beta$, the expectation
value of the corresponding operator is given by\footnote{Note that the action used here $S_{\beta=const}$ differs from
$S_{\alpha=const}$ by a boundary term.}
\begin{equation}
\label{scalarOb}
\langle {\cal O}_\beta \rangle = \frac{1}{\sqrt{\Omega}} \frac{\delta S_{\beta=const}}{\delta J_\beta} = -(\lambda_+ - \lambda_-) \alpha.
\end{equation}
From \eqref{asympt}, we can see that under a radial scaling $r \to
\lambda r$, $\alpha \to \lambda^{\lambda_-} \alpha$, while $\beta \to
\lambda^{\lambda_+} \beta$. This implies that the conformal dimensions
of the corresponding operators are indeed $\Delta_+ = \Delta( {\cal
  O}_\alpha) = \lambda_+$ and  $\Delta_- = \Delta( {\cal O}_\beta )
= \lambda_-$.  

Passing to more general boundary conditions will correspond to a
deformation away from exact conformal invariance, generating a
renormalization group flow. The deformation in the CFT action can be
computed by  using the bulk description to calculate the effect of the change in boundary
conditions on matrix elements and then  applying a Schwinger variational
principle. In the semi-classical limit, the change in boundary
conditions changes matrix elements in two ways: by changing the bulk
solution, and by changing the appropriate bulk  action; at each boundary condition, the bulk action must be chosen to give  a
well-defined variational principle. The result is that the deformation
of the field theory is determined by the corresponding potential in
\eqref{spot}. That is,
\begin{equation}
\label{sSFTa}
S^{FT} - S^{FT}_{\alpha=0} = \int dt \ {\cal W}_\alpha \big|_{\beta = \frac{1}{\lambda_+-\lambda_-} {\cal O}_\alpha} + {\cal O}(1/N),
\end{equation}
and
\begin{equation}
\label{sSFTb}
S^{FT} - S^{FT}_{\beta =0}  =  \int dt \ {\cal W}_\beta \big|_{\alpha = \frac{-1}{\lambda_+-\lambda_-} {\cal O}_\beta} + {\cal O}(1/N),
\end{equation}
where in each case, the argument of the potential is taken to be the
appropriate CFT operator.

 Let us consider a particularly simple example of a non-local boundary
condition, corresponding to $\gamma = -2$ in (\ref{gamma}).  The boundary condition imposes a
mode-dependent relation between the Fourier coefficients in the
expansion of the functions $\alpha(x)$ and $\beta(x)$ in Fourier modes
with respect to the {\it spatial} coordinates $\vec x$. For simplicity, we will
consider a linear relation,
\begin{equation} \label{modebc}
\alpha_k(t) = g \vec k^{-2} \beta_{\vec k}(t),
\end{equation}
for some constant $g$. In position space, this boundary condition would
correspond to $\alpha(x) = g \nabla^{-2} \beta(x)$, where $\nabla^2$
is the spatial Laplacian.

The boundary condition \eqref{modebc} thus corresponds to deforming the
$\Delta_+$ CFT by the relevant, non-local operator
\begin{equation}
{\cal W}_\alpha = \frac{g}{\lambda_+ - \lambda_-}  \int_{t=const} \sqrt{h} \  \tilde{\mathcal{O}} \nabla^{-2} \tilde{\mathcal{O}}.
\end{equation}
The $\Delta_+$ CFT corresponds to  $\alpha=0$ boundary conditions  in
the bulk\footnote{Note that our notation differs from that
of~\cite{Gubser:2002zh}.}. Since this is a relevant deformation, the
coupling $g$ has a positive conformal dimension. From the bulk point
of view, there is a natural prediction for the effect of this
perturbation: in the UV, $g \to 0$, and we will recover the $\Delta_+$
CFT $\alpha_k =0$. In the IR, $g \to \infty$, and we expect to find
the $\Delta_-$ CFT $\beta_k=0$. 

 The roles of the $\Delta_+$ and $\Delta_-$ CFTs are thus reversed
compared to an ultra-local deformation. Since~\cite{Gubser:2002zh} showed that $c_- >
c_+$, the central charge increases along our renormalization group flow.
We can see how this happens by following though the details of the
calculation of the (position-dependent) correction to the bulk vacuum energy as
in~\cite{Gubser:2002zh}. The analysis of the effect of the boundary
conditions on the propagator was carried out mode by mode
in~\cite{Gubser:2002zh}, so by substituting  $f = \vec k^2/g = r^2 \tilde k^2/g$  in their eq.
(35), we see immediately that
\begin{equation}
V(x,g) - V(x,\infty) = - \frac{1}{2^{d-2} \pi^{\frac{d}{2}}
  \Gamma(\frac{d}{2})} \int_0^\nu d\tilde \nu \frac{\tilde
  \nu}{\Gamma(\tilde \nu) \Gamma(1-\tilde \nu)} \int_0^\infty d\tilde
k \frac{\tilde k^d}{\tilde k + \tilde g} [ K_\nu (\tilde k)]^2,
\end{equation}
where $\nu = \sqrt{m^2 - m^2_{BF}}$, and
\begin{equation}
\tilde g = 2^{-2\nu} \frac{\Gamma(1-\nu)}{\Gamma(1+\nu)} g \left( \frac{1}{r} \right) ^{2(1-\nu)}
\end{equation}
contains all of the position dependence. The integral on the RHS is
negative, and is a monotonic function of $\tilde g$. Hence, we can see that
$V(x,g)$ is a monotonically increasing function of $\tilde g$, reaching its
maximum value at $\tilde g=\infty$. That is, the correction to the
cosmological constant increases towards the interior of the spacetime,
corresponding to an increasing central charge in the IR.

\section{Discussion}
\label{disc}

Our comments above considered non-local 
boundary potentials ${\cal W}_\alpha$, ${\cal W}_\beta$ in the context of AdS/CFT.
For the particular boundary condition $\alpha = g \nabla^{-2} \beta$, where $\nabla^2$ is the spatial Laplacian, 
we computed the position-dependent
one-loop vacuum energy following \cite{Gubser:2002zh} and found that 
the corresponding renormalization group flow runs from the $\Delta_+$ CFT in the UV to the $\Delta_-$ CFT in the IR.
As a result, the effective central charge {\it increases} along this flow, and the endpoints of the flow are interchanged relative to flow associated with the more familiar ultra-local boundary conditions.  This is consistent because the UV description of
our flow is given by a non-local perturbation of the $\Delta_+$ CFT.

The above example is a special case of the boundary condition (\ref{gamma}) with $\gamma =-2$. 
Let us briefly consider the more general case, but restrict ourselves to $F_\gamma$ such that either $F_\gamma$ or $[F_\gamma]^{-1}$ is local.  This requires $\gamma \in {\mathbb Z}$.  The case $\gamma=0$ is familiar, and for the remaining cases only one of ${\cal W}_\alpha, {\cal W}_\beta$ can be local.    As usual, we restrict to the
mass range (\ref{mrange}).
For $\gamma \ge 2$, the local potential is ${\cal W}_\beta$ and is irrelevant.  Here the effective central charge decreases along the RG flow.
For $\gamma \le -1$ the local potential  is ${\cal W}_\alpha$ and is again irrelevant.  Such cases are like the example of section~\ref{vace}, and the effective central charge increases along the RG flow.
The remaining case $\gamma =1$ is more interesting, however.  Here the local potential is ${\cal W}_\beta$.
For small mass ($m^2 - m^2_{BF} < 1/4$), 
this potential is irrelevant.  However, it becomes 
marginal when $m^2 - m^2_{BF} = 1/4$, and then relevant for larger masses. 
We see that when both $F_1 = v^i \nabla_i$ for some vector field $v^i$ and $m^2 - m^2_{BF} \ge 1/4$, the boundary dual admits a UV-complete local description. This suggests that the dynamics of bulk fields should be fully local on conformally compactified AdS.  While it is not apparent to us
what makes such cases special from the bulk perspective, it is clear that further study is warranted.
The case $m^2 -m^2_{BF}=1/4$ is particularly interesting, as it arises in the usual maximal gauged supergravity theory on AdS${}_4$ \cite{dWN1,dWN2}, which can in fact be consistently truncated to a theory involving only gravity and this scalar \cite{DL}.   

\medskip

\textbf{Acknowledgements:} We would like to thank David
Berenstein,  Andreas Karch,  Joe Polchinski, and Matt Strassler
for interesting discussions related to this work.  This research was supported in part by the National
Science Foundation under Grants No. PHY99-07949, No. PHY03-54978,
by funds from the University of California, and by the EPSRC.

\bibliographystyle{utphys}

\bibliography{nonlocal4}

\end{document}